\documentclass[final,1p,times,authoryear]{elsarticle}

\usepackage{amssymb}
\usepackage{doi}
\usepackage{tabularx}
\usepackage{bbding}
\usepackage{multirow}
\usepackage{amsmath}
\usepackage{booktabs}
\usepackage{caption}
\usepackage{adjustbox}
\usepackage{float}
\biboptions{numbers,sort&compress}
\usepackage{color}
\usepackage{stfloats}
\usepackage{subcaption}
\usepackage{threeparttable}
\begin{document}

\begin{frontmatter}



\title{Attack Smarter: Attention-Driven Fine-Grained Webpage Fingerprinting Attacks}


\author[first]{Yali Yuan}
\ead{yaliyuan@seu.edu.cn}
\author[first]{Weiyi Zou}
\ead{zouweiyi@seu.edu.cn}
\author[first]{Guang Cheng \corref{cor1}}
\ead{chengguang@seu.edu.cn}
\affiliation[first]{organization={School of Cyber Science and Engineering},
            addressline={Southeast University}, 
            city={Nanjing},
            postcode={211189}, 
            state={Jiangsu Province},
            country={China}}

\cortext[cor1]{Corresponding author at: School of Cyber Science and Engineering, Southeast University, Nanjing, 211189, Jiangsu Province, China}

\begin{abstract}
Website Fingerprinting (WF) attacks aim to infer which websites a user is visiting by analyzing traffic patterns, thereby compromising user anonymity. Although this technique has been demonstrated to be effective in controlled experimental environments, it remains largely limited to small-scale scenarios, typically restricted to recognizing website homepages. In practical settings, however, users frequently access multiple subpages in rapid succession, often before previous content fully loads. WebPage Fingerprinting (WPF) generalizes the WF framework to large-scale environments by modeling subpages of the same site as distinct classes. These pages often share similar page elements, resulting in lower inter-class variance in traffic features. Furthermore, we consider multi-tab browsing scenarios, in which a single trace encompasses multiple categories of webpages. This leads to overlapping traffic segments, and similar features may appear in different positions within the traffic, thereby increasing the difficulty of classification. To address these challenges, we propose an attention-driven fine-grained WPF attack, named ADWPF. Specifically, during the training phase, we apply targeted augmentation to salient regions of the traffic based on attention maps, including attention cropping and attention masking. ADWPF then extracts low-dimensional features from both the original and augmented traffic and applies self-attention modules to capture the global contextual patterns of the trace. Finally, to handle the multi-tab scenario, we employ the residual attention to generate class-specific representations of webpages occurring at different temporal positions. Extensive experiments demonstrate that the proposed method consistently surpasses state-of-the-art baselines across datasets of different scales.
\end{abstract}

\begin{keyword}
Webpage fingerprinting \sep Tor \sep Data augmentation \sep Attention \sep Multi-label learning

\end{keyword}

\end{frontmatter}



\section{Introduction}
\label{Introduction}

Tor \citep{dingledine2004tor}, one of the most prominent and widely employed anonymous communication systems, is used to safeguard user browsing privacy. 
It conceals the path between users and visited websites through a three-hop proxy, ensuring that each relay only has access to the IP addresses of its predecessor and successor.
Although packet payloads are protected by multilayer encryption that prevents the extraction of content, traffic patterns, such as packet direction and timestamps, can still expose browsing activities.  
Inferring visited websites through the analysis of Tor traffic is known as website fingerprinting (WF) attacks.
Various approaches have been proposed to improve the accuracy of WF attacks, including the incorporation of time-based features \citep{bhat2018var,rahman2019tik}, the development of novel model architectures \citep{guan2021bapm,jin2023transformer}, and the adoption of advanced training strategies \citep{sirinam2019triplet,bahramali2023realistic}.
Notably, Tik-Tok \citep{rahman2019tik} achieves an impressive 98.4\% accuracy on 95 websites in the closed-world setting without defenses. 
However, these methods predominantly concentrate on a limited set of monitored websites and often overlook a common user behavior: after visiting a homepage, users frequently navigate to various subpages within the same site. 
In distinguishing between these subpages, conventional models often fail to capture discriminative features, leading to a substantial decline in performance.

Current WPF techniques \citep{shen2019webpage, zhang2019deep, shen2021efficient, lu2021gap} are predominantly evaluated under single-tab scenarios, with datasets that typically contain around 100 classes. This experimental setup does not accurately reflect their performance in real-world environments. 
This raises several critical questions: Can these models maintain strong performance when the number of webpage classes increases substantially? Moreover, in situations where users open multiple pages in rapid succession, does the presence of obfuscated segments within a trace impair the predictive capabilities of these models?
Recently, Oscar \citep{zhao2024towards} addressed the aforementioned challenges by evaluating its model on a multi-tab dataset comprising 1,000 webpage classes. However, despite this progress, the approach still exhibits several limitations. 
Oscar adopts a metric learning framework that maps traffic direction into a feature space, which necessitates the storage of both model parameters and proxy nodes after training. 
During the inference phase, it must compute nearest-neighbor distances not only between samples and proxies but also among different samples, leading to significantly higher time overhead compared to end-to-end architectures. 
Furthermore, Oscar applies random inter-class and intra-class augmentation techniques to increase dataset diversity. However, such random augmentation strategies may introduce uncontrolled noise into the data distribution.

In this paper, we propose an end-to-end WPF attack, called ADWPF.
To enable the model to distinguish webpages based on subtle differences, we propose an attention-based data augmentation strategy \citep{hu2019see} that guides the model, in a weakly supervised manner, to attend to discriminative segments while gradually learning from previously ignored regions. 
Additionally, we leverage a self-attention mechanism \citep{vaswani2017attention} to enable the model to capture global patterns and extract richer semantic features. 
Since a single trace may contain multiple webpages, conventional classifiers struggle to perceive position-specific discriminative features. 
To address this, residual attention \citep{zhu2021residual} effectively integrates class-agnostic global features with class-specific local features, enabling strong performance in multi-tab recognition tasks without introducing excessive model complexity.

Specifically, the model utilizes a Convolutional Neural Network (CNN) to extract local features from traffic and uses multi-head self-attention from the Transformer architecture to construct global representations. 
During training, attention maps from the final CNN layer are leveraged to identify salient regions within the traces - regions with higher response values are more likely to be selected as guidance for data augmentation. 
In the attention cropping, these salient segments are retained, encouraging the model to perform classification based solely on these discriminative features.
In contrast, attention masking discards the salient segments, forcing the model to extract features from the less distinctive portions of the traffic.
Given the variable number of webpages within a trace, we introduce a residual attention mechanism at the prediction head to more effectively align the extracted features with their corresponding labels. 
By integrating class-specific and class-agnostic representations, the model can capture discriminative patterns distributed across varying positions in the traffic. 
Throughout both training and inference, the attention mechanism significantly reduces the impact of background noise and ambiguous segments, enabling the model to focus on subtle differences in the feature map and enhance its ability to perform the effective WPF attack.

We evaluated ADWPF's performance under both closed-world and open-world settings. 
The closed-world dataset comprises 1,000 monitored webpages, and the number of webpages per trace is dynamic, thereby preventing the model from exploiting prior knowledge about the number of pages. 
Compared to existing state-of-the-art approaches, ADWPF achieves a mean Average Precision (mAP) of 50.54\% and a Recall@5 of 63.85\%, surpassing the best existing method by 13.49\% and 10.08\%, respectively. 
Even as the number of webpage classes scales, ADWPF consistently maintains a Recall@5 above 60\%.

In summary, our attention-driven approach demonstrates strong effectiveness, and the contributions of this work are summarized in four key contributions.

\begin{itemize}
    \item We propose ADWPF, an attention-driven, fine-grained WPF attack. ADWPF achieves accurate identification even in large-scale, multi-tab traffic scenarios with unknown numbers and types of webpages.
    
    \item We incorporate attention maps for data augmentation, in contrast to conventional methods that perform direct transformations on raw traffic inputs. Instead, ADWPF integrates augmentation into the training pipeline by cropping and masking salient traffic segments. This strategy guides the model to focus on local discriminative features and discover previously ignored patterns.
    
    \item We employ a residual attention mechanism to improve the model's performance in multi-tab prediction scenarios. By fusing global and local features, the model can adaptively capture class-specific representations, even when multiple webpages are present within a single trace.

    \item We benchmark ADWPF against state-of-the-art models on a dataset of 1,000 monitored webpages and validate its effectiveness across both closed-world and open-world settings, demonstrating that attention mechanisms can significantly improve performance.
\end{itemize}

The remainder of this paper is organized as follows: In Section \ref{Background and related work}, we review related work on WF and WPF attacks, as well as the technical details.
Section \ref{Threat model} formulates the threat model for multi-tab WPF. 
Section \ref{Methodology} describes the proposed method in detail, including the implementation of each module in the model. 
Section \ref{Experiment} outlines the experimental setup and reports the corresponding results. 
In Section \ref{Discussion}, we discuss the contributions of this work and suggest possible avenues for future research and improvement. 
Finally, Section \ref{Conclusion} provides concluding remarks.

\section{Background and related work}
\label{Background and related work}

WF attacks can be formulated as a classification task, where encrypted traffic serves as the input samples and website categories as the labels. 
The attack process involves extracting traffic features and feeding them into a model to obtain predictions. 
This section provides a review of WF and WPF attacks in both single-tab and multi-tab settings, as summarized in Table \ref{wftable}. It also provides background knowledge on fine-grained classification and multi-label classification.

\subsection{Single-tab WF attacks}

Early approaches to single-tab WF primarily relied on manually engineered features, followed by classification using traditional machine learning algorithms. 
\cite{herrmann2009website} were the first to apply Bayesian classification to WF attacks on Tor, but due to the use of only packet length as the feature, the approach resulted in a low classification accuracy of 3\% on a sample of 775 sites.
Subsequently, researchers explored a wide range of traffic features in the domain of WF attacks.
\cite{panchenko2011website} improved the attack accuracy to 55\% by using a Support Vector Machine (SVM) classifier trained on statistical features of traffic, including volume, time, and direction.
\cite{wang2013improved} used the direction of Tor cells as features and further enhanced the effectiveness of WF attacks by removing noise cells and introducing a novel distance metric.
Moreover, \cite{wang2014effective} engineered nearly 4,000 features for similarity measurement, achieving 85\% true positive rate and 0.6\% false positive rate using the k-Nearest Neighbors (k-NN) algorithm on a dataset of 100 monitored websites and 5,000 non-monitored websites.
\cite{panchenko2016website} sampled features from the cumulative sum of packet lengths and employed SVM for classification, while \cite{hayes2016k} leveraged random forests to extract raw traffic features by treating leaf nodes as features for subsequent k-NN classification.
Both approaches achieved over 90\% accuracy using 90 traffic samples per class. 
However, feature engineering in traditional WF attacks is labor-intensive and time-consuming, relying heavily on domain expertise, which hinders the adaptability of models in dynamic network environments.

The emergence of deep learning techniques facilitated the development of end-to-end WF attacks,, which reduce reliance on manual feature engineering while enhancing classification accuracy. \cite{rimmer2017automated} utilized Convolutional Neural Network (CNN), Long Short-Term Memory (LSTM), and Stacked Denoising Autoencoder (SDAE) to perform attacks across 900 websites. Their CNN-based attack achieved an accuracy of 96.3\%, highlighting the potential of deep learning models for WF tasks.
Subsequently, Deep Fingerprinting (DF) \citep{sirinam2018deep} leveraged the capabilities of CNN by designing a deeper network architecture with regularization techniques to prevent overfitting, achieving superior performance over traditional WF attacks in both closed-world settings and defense scenarios. 
Later, unlike these approaches that relied solely on direction, Tik-Tok \citep{rahman2019tik} built upon the DF model by combining direction and timestamps through element-wise multiplication to generate more informative features.
Similarly, Var-CNN \citep{bhat2018var} separately extracted direction and inter-packet time features for training. When trained with 40 samples per website, it achieved a classification accuracy of 92.4\%. The incorporation of direction and time features further enhanced the model’s performance.

In addition, to facilitate domain adaptation, Triplet Fingerprinting (TF) \citep{sirinam2019triplet} introduced few-shot learning into the field of WF attacks. It utilized a triplet network to pull samples of the same class closer together while pushing samples of different classes apart in the feature space. NetCLR \citep{bahramali2023realistic} leveraged contrastive learning during pretraining to mitigate dependence on labeled data, and subsequently achieved 80\% accuracy in the closed-world setting through fine-tuning. All of the aforementioned approaches are limited to website-level identification. In contrast, WPF, which targets individual pages within a website where feature differences are more subtle, has recently attracted increasing attention.

\subsection{Single-tab WPF attacks}

Since subpages within a website often share highly similar page elements, WPF necessitates the identification of fine-grained traffic features. 
Similar to WF, WPF methods include both traditional machine learning and deep learning approaches. 
For instance, \cite{shen2019webpage} applied the k-NN algorithm for early-stage WPF using only the first 100 packets of traffic, achieving 91.6\% accuracy. 
\cite{zhang2019deep} leveraged the local request and response sequence (LRRS) as features combined with a Random Forest classifier. 
Among deep learning methods, BurNet \citep{shen2021efficient} employed CNN to extract burst-level features, while GAP-WF \citep{lu2021gap} developed flow-level graphs to learn both intra-flow and inter-flow representations, leveraging Graph Neural Network (GNN) for feature extraction. 
These methods identify webpages by extracting both fine-grained and coarse-grained features, demonstrating substantial performance. 
However, they are constrained to single-tab scenarios, posing challenges for extension to more realistic multi-tab settings.

\subsection{Multi-tab WF and WPF attacks}

When users open multiple pages simultaneously,  asynchronous page load speeds may cause one page to begin loading before the previous one has finished,  resulting in certain segments within a trace containing features from multiple webpage categories. 
This makes it difficult to extract clean and distinct traffic patterns. 
To mitigate the interference caused by such overlapping segments, prior approaches primarily relied on splitting methods to isolate single-tab traffic features before classification.
\cite{wang2016realistically} initially introduced two segmentation strategies: time-based splitting and classification-based splitting. 
By converting multi-tab traffic into isolated single-tab segments, they then applied k-NN for classification. 
\cite{xu2018multi} employed a BalanceCascade-XGBoost method to identify boundaries between consecutive pages, which enhanced the classification accuracy of the preceding page. \cite{cui2020more} investigated leveraging a limited set of packets surrounding the boundary, combined with CNN and LSTM for prediction.
Although these methods successfully extracted distinguishable traffic features through splitting, their pipelines are complex, and the segmentation process inevitably results in the loss of some useful information.

To reduce the dependence on clean traffic segments and better exploit the overlapping portions of traffic, BAPM \citep{guan2021bapm} adopted a multi-head attention mechanism to adaptively analyze multi-tab traffic and perform end-to-end prediction. 
However, it necessitated prior knowledge of the number of pages involved. TMWF \citep{jin2023transformer} introduced the concept of object detection by assigning a tab query to each page to retrieve class-specific features.  It further eliminated the need for explicit page count information and proposes a new validation method to address redundant predictions.
ARES \citep{deng2023robust} reformulated multi-tab WF attacks as a multi-label classification task, thereby removing the need for prior knowledge of page count. 
It extracted local features using CNN and applied a Top-k self-attention mechanism to model global dependencies while reducing the influence of noise, achieving robust performance on large-scale, real-world multi-tab datasets.
To enable multi-tab WPF, Oscar \citep{zhao2024towards} conducted experiments on 1,000 webpage classes, where the number of pages per traffic instance was entirely unknown. 
It employed a metric learning approach to cluster traffic from the same page and separated traffic from different pages, making it the first large-scale, fine-grained, multi-tab WPF attack capable of accurate identification. 
However, Oscar requires additional storage for proxy classes and incurs substantial computational overhead during inference due to the need to compute distances between samples and proxies. 
To address these limitations, we propose a model capable of performing fine-grained, end-to-end prediction without reliance on proxy-based and sample-based comparisons.

\begin{table}[h!]
\centering
\caption{A summary of representative WF and WPF attacks.}
\begin{adjustbox}{width=\linewidth}
\label{wftable}
\begin{threeparttable}
\begin{tabular}{ccccccccc}
\toprule

\multirow{2}{*}{\textbf{Research}} & \multicolumn{3}{c}{\textbf{Traffic features}} & \multirow{2}{*}{\textbf{Multi-tab}} & \multirow{2}{*}{\textbf{Dynamic prediction}} \tnote{1} & \multirow{2}{*}{\textbf{Target}} & \multirow{2}{*}{\textbf{Data augmentation}} & \multirow{2}{*}{\textbf{Method}} \\ 
\cmidrule(lr){2-4}
& \textbf{Time} & \textbf{Direction} &  \textbf{Length}          &     \\ 
\midrule

AWF \citep{rimmer2017automated} &    \XSolidBrush      & \CheckmarkBold & \XSolidBrush &  \XSolidBrush  &  \XSolidBrush & Website & No  & SDAE, CNN, LSTM \\

DF \citep{sirinam2018deep} &    \XSolidBrush      & \CheckmarkBold & \XSolidBrush & \XSolidBrush   & \XSolidBrush  & Website& No & CNN \\

Var-CNN \citep{bhat2018var} &      \CheckmarkBold   & \CheckmarkBold & \XSolidBrush &  \XSolidBrush  & \XSolidBrush  & Website& No & ResNet-18, Dilated causal convolutions \\

Tik-Tok \citep{rahman2019tik} &    \CheckmarkBold    & \CheckmarkBold & \XSolidBrush &  \XSolidBrush & \XSolidBrush  & Website & No & CNN \\

TF \citep{sirinam2019triplet} &   \XSolidBrush     & \CheckmarkBold & \XSolidBrush &  \XSolidBrush   & \XSolidBrush  & Website & No & CNN,Triplet networks  \\

NetCLR \citep{bahramali2023realistic} &   \XSolidBrush     & \CheckmarkBold & \XSolidBrush &  \XSolidBrush   & \XSolidBrush  & Website & Random augmentation & CNN, Self-supervised learning\\

BurNet \citep{shen2021efficient} &   \XSolidBrush     & \XSolidBrush & \CheckmarkBold &  \XSolidBrush   & \XSolidBrush  & Webpage & No & CNN \\

GAP-WF \citep{lu2021gap} &   \CheckmarkBold     & \CheckmarkBold & \CheckmarkBold &  \XSolidBrush   & \XSolidBrush  & Webpage & No & GNN \\

BAPM \citep{guan2021bapm}         &      \XSolidBrush         &   \CheckmarkBold & \XSolidBrush   & \CheckmarkBold  & \XSolidBrush &  Website & No  & CNN, Multi-head self-attention \\
TMWF \citep{jin2023transformer}         &     \XSolidBrush          &    \CheckmarkBold  & \XSolidBrush    &  \CheckmarkBold  &  \CheckmarkBold & Website & No & CNN, Transformer          \\
ARES \citep{deng2023robust}         & \XSolidBrush &    \CheckmarkBold & \XSolidBrush  &   \CheckmarkBold  & \CheckmarkBold    &  Website  & No     & CNN, Top-k multi-head self-attention          \\
Oscar \citep{zhao2024towards}         & \XSolidBrush &    \CheckmarkBold & \XSolidBrush  &   \CheckmarkBold  & \CheckmarkBold    &  Webpage  & Random augmentation     & CNN, k-NN, Metric learning          \\
ADWPF         & \XSolidBrush &    \CheckmarkBold & \XSolidBrush  &   \CheckmarkBold  & \CheckmarkBold    &  Webpage & Attention augmentation       & CNN, Multi-head self-attention          \\

\bottomrule
\end{tabular}

\begin{tablenotes}
    \item[1] In multi-tab traffic, prediction is performed without requiring prior knowledge of the number of pages. 
\end{tablenotes}

\end{threeparttable}
\end{adjustbox}
\end{table}

\subsection{Fine-grained classification}

Fine-grained image recognition methods are generally categorized into localization methods \citep{zhang2014part,lin2015deep,fu2017look,wang2019weakly,hu2019see} and feature-encoding methods \citep{yu2018hierarchical,zheng2019learning}. The former focuses on identifying discriminative regions and leveraging their subtle differences for classification, while the latter enhances feature representations through high-order feature interactions. Our work primarily explores the localization methods.
Early methods \citep{zhang2014part,lin2015deep} relied on additional part annotations to localize key regions, which were highly labor-intensive. Consequently, subsequent research \citep{fu2017look,wang2019weakly,hu2019see} shifted toward weakly supervised techniques to identify local informative regions without requiring explicit annotations. Inspired by the work of \citep{hu2019see}, which employed attention maps for image cropping and dropping to encourage the model to focus on more discriminative regions, we adopt a similar strategy for traffic data augmentation. The detailed procedure is described in the subsequent sections.

\subsection{Multi-label classification}
Multi-label classification allows the prediction of a set of labels from a given sample. In computer vision, many existing approaches enhance prediction performance by leveraging semantic relationships \citep{chen2019learning} or designing novel loss functions \citep{ridnik2021asymmetric}. In contrast, residual attention \citep{zhu2021residual} introduced an advanced classification head capable of identifying the corresponding positions of different labels within a trace, without significantly increasing the number of parameters. This makes it particularly well-suited for multi-tab WPF attacks.

\section{Threat model}
\label{Threat model}

\begin{figure}[!ht]
	\centering 
	\includegraphics[width=0.8\linewidth]{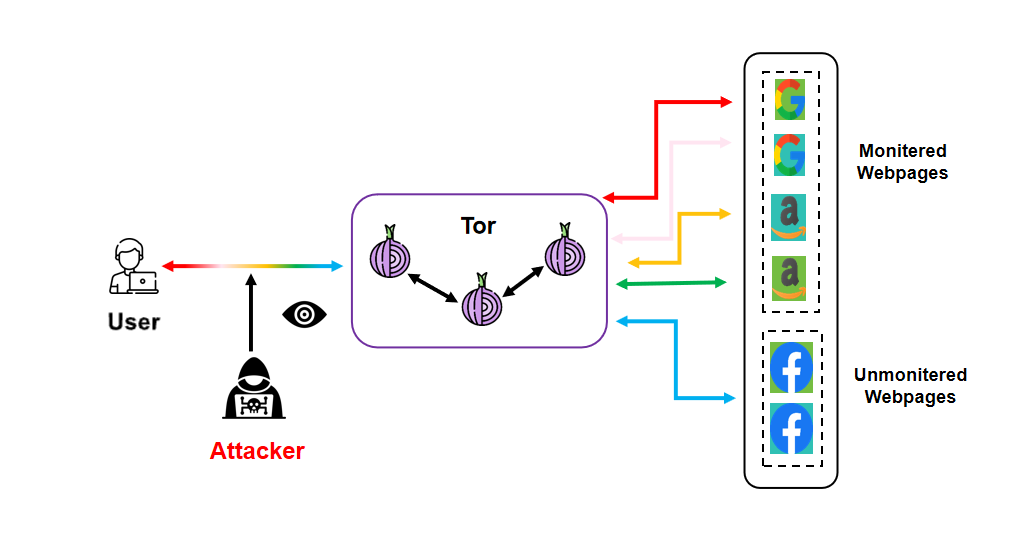}	
	\caption{Threat model of multi-tab WPF attacks. Users may concurrently access webpages from both a monitored set and an unmonitored set. For monitored pages, the attacker's objective is to perform fine-grained classification, accurately identifying individual subpage categories. In contrast, for unmonitored pages, the goal is to detect and classify them simply as unmonitored, without attempting to determine their exact category.} 
	\label{threat}%
\end{figure}

The threat model of ADWPF is illustrated in Fig.\ref{threat}. 
WF attacks can potentially occur in various locations, such as local system administrators, Internet Service Providers (ISPs), users' networks, or Autonomous Systems (AS) between users and Tor entry nodes. The adversary passively collects traffic data by monitoring the communication between the user's Tor client and the Tor network entry node, without modifying, deleting, or injecting any packets. 
The attacker extracts features from captured traffic that are uniquely associated with each webpage. These features generally fall into two categories: statistical features and per-packet feature sequences. Statistical features are computed over a trace, such as the total number of packets or overall transmission time, whereas per-packet features include packet size, Tor cell direction, and timestamp. By aggregating these features and inputting them into a classifier for training, the attacker can then deploy the model in real-world environments to infer users’ visited websites from newly captured traffic.
Unlike many previous assumptions \citep{sirinam2018deep, sirinam2019triplet,bahramali2023realistic} where users are restricted to accessing only the homepage and wait for one page to fully load before proceeding, our threat model considers a more realistic scenario in which users randomly click on multiple links within a website simultaneously. This renders it infeasible for the attacker to know in advance the number or types of pages visited. Furthermore, since subpages are significantly more numerous than websites, the monitored set encompasses thousands of distinct webpage categories.

It is impractical to collect Tor traffic data for all websites on the Internet. Consequently, many existing studies restrict their focus to a small set of monitored websites and evaluate model performance within this constrained scope, a setting commonly referred to as the closed-world. However, \cite{juarez2014critical} pointed out the limitations of this assumption. In response, subsequent research \citep{hayes2016k,panchenko2016website,sirinam2018deep,sirinam2019triplet} introduced traffic from unmonitored websites into the dataset, aiming to provide a more realistic assessment of model performance, known as the open-world. In such scenarios, users may simultaneously visit both monitored and unmonitored webpages, while the attacker has access to only a limited amount of data from unmonitored websites for training.

\section{Methodology}
\label{Methodology}

\subsection{Preliminary}

In the context of multi-tab WPF, a single traffic trace may correspond to multiple webpage categories. Unlike single-tab scenarios, where the output is a single class index, the label is represented as a one-hot vector in this case. Given a dataset $\mathcal{D}=\{(x_i, Y_i)\}_{i=1}^{N}$ of size $N$, each traffic instance $x_i\in \mathbb{R}^l$ is preprocessed to a fixed length $L$, and its corresponding label is represented as a binary vector $Y_i=[Y_{i,1}, Y_{i,2},\ldots, Y_{i,W_n}]\in\{0,1\}^{W_n}$, where $W_n$ is the total number of webpage categories. $Y_{i,j}=1$ indicates that the $j$-th webpage category is present in  $x_i$; otherwise, $Y_{i,j}=0$.

\subsection{Overview}

\begin{figure}
	\centering 
	\includegraphics[width=0.99\linewidth]{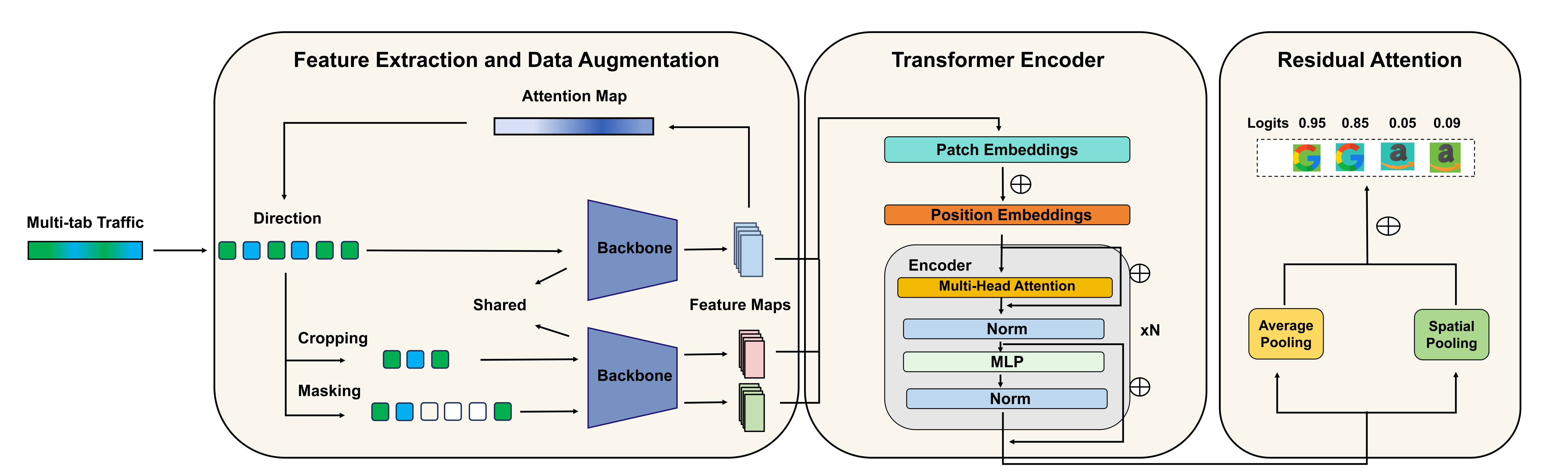}	
	\caption{The architecture of ADWPF consists of four main modules: Feature Extraction, Data Augmentation, Transformer Encoder, and Residual Attention.}
	\label{model}%
\end{figure}

ADWPF is an end-to-end, attention-driven framework for WPF attacks. It integrates a CNN and Transformer-based architecture to capture both local and global traffic patterns. The framework utilizes coarse-grained labels and generates attention maps from convolutional layers to identify discriminative regions within traffic, which are then used for data augmentation. This process does not rely on expert knowledge to manually select augmentation strategies. Instead, ADWPF autonomously learns which regions are most informative for classification. At the classification head, a lightweight, training-free residual attention mechanism enables the model to associate different segments of the sequence with distinct webpage categories. These components allow ADWPF to distinguish subtle differences across multiple overlapping webpages simultaneously.

The architecture of ADWPF is illustrated in Fig.\ref{model} and comprises four modules. First, traffic direction is extracted as the input feature and fed into a feature extractor, which produces both feature maps and attention maps. This extractor consists of a multi-layer CNN architecture designed to capture low-level traffic representations.
Next, a randomly selected attention map guides two complementary data augmentation operations: attention cropping and attention masking. Cropping retains only the most salient regions of the sequence, helping the model focus on discriminative patterns. Masking, on the other hand, preserves the less salient regions, encouraging the model to explore overlooked yet potentially informative features. These two augmented sequences are passed through the shared feature extractor to produce enriched feature representations.
The Transformer Encoder module applies multi-head self-attention to model global dependencies across the feature maps. This enables the model to dynamically aggregate information from different positions and enhances its ability to recognize complex patterns.
Finally, the Residual Attention module generates classification outputs. It combines spatial pooling, which captures class-specific activations across sequence positions, with global average pooling, which captures class-agnostic features. The outputs of these two pooling mechanisms are summed to compute the final multi-label prediction probabilities.

\subsection{Feature extraction}

\begin{figure}[!ht]
	\centering
	\includegraphics[width=0.8\linewidth]{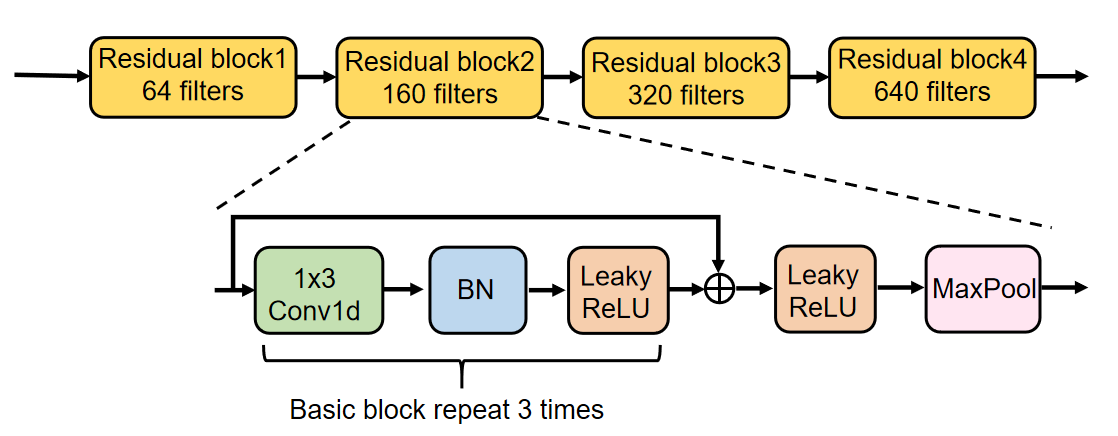}	
	\caption{The architecture of 1D ResNet-12.} 
	\label{resnet}%
\end{figure}

Deeper and wider feature extractors effectively reduce intra-class variance and improve the model's capacity to capture complex patterns.  While many existing studies adopt DF \citep{sirinam2018deep,rahman2019tik,bahramali2023realistic} as the backbone, our experiments demonstrate that ResNet \citep{he2016deep} exhibits superior generalization to unseen websites when used as the backbone. To tailor ResNet for sequential data, we adapt the standard ResNet-12 architecture \citep{tian2020rethinking}, as illustrated in Fig.\ref{resnet}. The modified model comprises four residual blocks. Each residual block contains three basic blocks, with each block consisting of a convolutional layer, a batch normalization layer, and a LeakyReLU activation function. To stabilize training, a residual connection is established by adding the block's input to its output. Finally, a max pooling layer is applied at the end of each residual block to perform downsampling.

For a trace $x_{i}=\{d_1,d_2,...,d_L\}$ of length $L$, each element $d_i$ represents the direction of a packet, where +1 indicates an outgoing packet and -1 indicates an incoming packet. 
Since traces vary in length, zero-padding is applied to the end of each sequence to standardize them to a fixed length $L$, enabling batch processing by the feature extractor. 
To map the input sequence into a low-dimensional representation, we define a feature extraction model $f_{\boldsymbol{\theta}}:\mathbb{R}^{L\times 1}\to\mathbb{R}^{M\times C}$ parameterized by learnable weights $\theta$. The output $z\in \mathbb{R}^{M\times C}$ is a feature map, where $C$ is the number of channels and $M$ is the length. The computation of the feature map is formally expressed as:

\begin{equation}
    z=f_{\theta}(x),
\end{equation}

In addition, we obtain attention maps $a\in \mathbb{R}^{M \times C^{\prime}}$ from the final convolutional layer, which are used in the subsequent data augmentation stage. Here, $C^{\prime}$ denotes the number of attention maps,  each corresponding to a distinct learned focus region within the input sequence.

\subsection{Data augmentation}

\begin{figure}[!ht]
	\centering
	\includegraphics[width=0.8\linewidth]{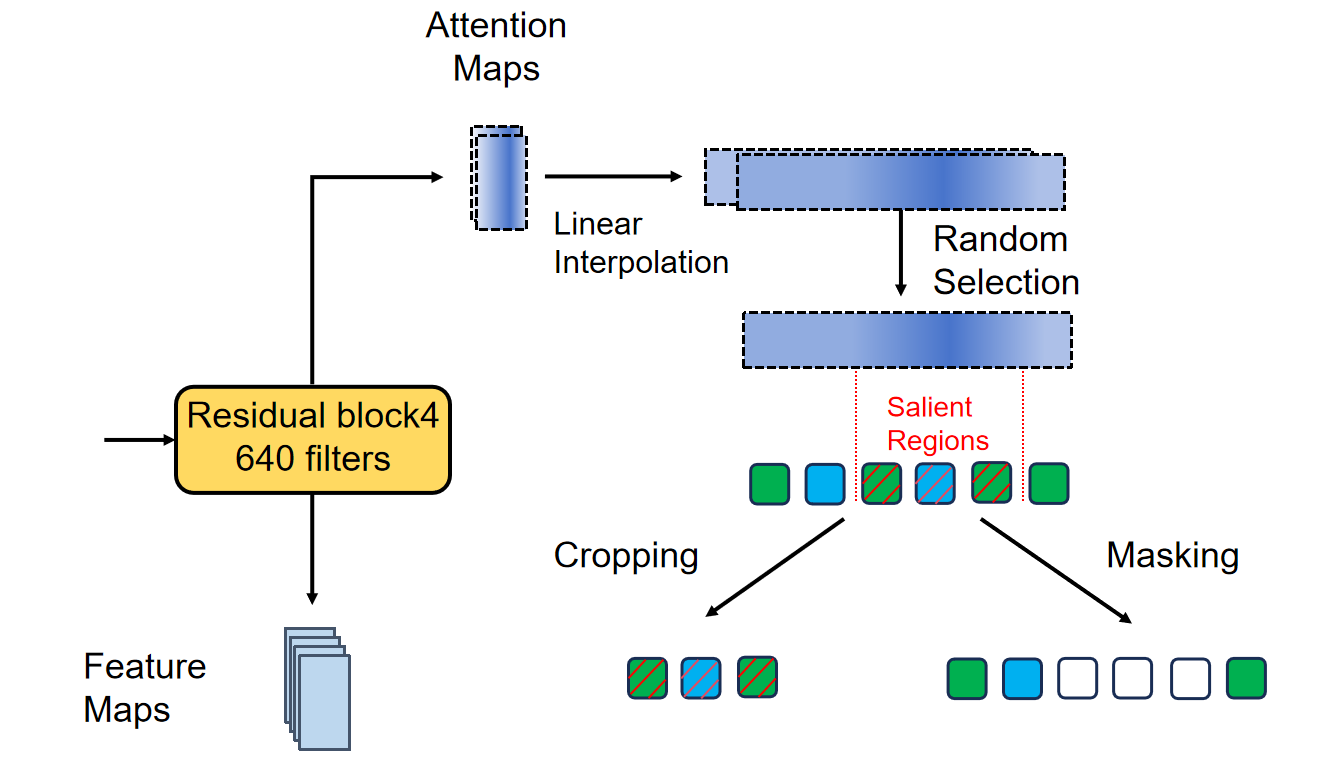}	
	\caption{The data augmentation process includes attention cropping and attention masking.} 
	\label{Augmentation}%
\end{figure}

Since there are no explicit annotations indicating which segments of a trace correspond to specific webpages, the model must implicitly learn the positional distribution associated with each label. The attention map $a$ serves as the classification guide, highlighting regions considered informative. Compared to random data augmentation, attention-guided augmentation reduces the likelihood of introducing noise and enhances the effectiveness of the learning process by focusing on discriminative segments. 

The data augmentation process is illustrated in Fig.\ref{Augmentation}. To ensure alignment between the attention maps and the original input sequence, the attention maps $a$ are first upsampled to the original sequence length $L$ using linear interpolation, yielding $a^{*}\in \mathbb{R}^{L \times C^{\prime}}$.
One attention map $a^{*}_k\in \mathbb{R}^{L}$ is then randomly selected from $a^{*}$ to guide the augmentation. For $a^{*}_k$,  the importance of each position is determined based on its activation value. Given a predefined threshold $\varphi_c$, Each position $a^{*}_k(i)$ is assigned a binary value $s_c(i) \in\{0,1\}^{L}$: if $a_k^*(i)>\varphi_c$, then $s_c(i)=1$, and $s_c(i)=0$, otherwise. This index sequence $s_c$ serves to identify and retain important segments during the cropping operation. The cropping indicator is defined as:

\begin{equation}
s_c(i)=
\begin{cases}
1, & \mathrm{if~}a^{*}_k(i)> \varphi_c \\
0, & \text{otherwise.}
\end{cases} \quad\mathrm{for}i=1,2,\ldots,L,
\end{equation}

Similarly, to encourage the model to discover potentially discriminative features in less salient regions, we construct a binary mask sequence $s_m \in\{0,1\}^{L}$ to indicate the areas to be masked. Given a predefined threshold $\varphi_m$, each element of $s_m$ is set to 0 if the corresponding value in $a^{*}_k$ exceeds $\varphi_m$; otherwise, it is set to 1. The masking indicator is defined as:

\begin{equation}
s_m(i)=
\begin{cases}
0, & \mathrm{if~}a^{*}_k(i)> \varphi_m \\
1, & \text{otherwise.}
\end{cases} \quad\mathrm{for}i=1,2,\ldots,L,
\end{equation}

By performing element-wise multiplication between $s_c$ and $s_m$ with the original input sequence $x$, we obtain the augmented sequences $x^\mathrm{crop}$ and $x^\mathrm{mask}$. This process can be formally expressed as follows:

\begin{equation}
x_i^\mathrm{crop}=x_i \cdot s_c(i),\quad\mathrm{for} i=1,2,\ldots,L,
\end{equation}

\begin{equation}
x_i^\mathrm{mask}=x_i \cdot s_m(i),\quad\mathrm{for} i=1,2,\ldots,L,
\end{equation}

$x^\mathrm{crop}$ and $x^\mathrm{mask}$ are passed into the feature extractor with the same weights to obtain a new feature map. For convenience in subsequent steps, both the original and augmented feature maps are denoted as $z$.

\subsection{Transformer encoder}

In both the computer vision \citep{dosovitskiy2020image} and natural language processing \citep{devlin2019bert} domains, numerous task-specific models are developed based on the Transformer \citep{vaswani2017attention}. Its primary advantage lies in the self-attention mechanism, which enables global modeling over sequences by capturing dependencies between different segments. This makes the Transformer particularly well-suited for processing direction sequences in Tor traffic. However, directly processing the entire raw sequence with the Transformer can be computationally expensive and may result in poor convergence during training. To address this, we first use a CNN-based module to extract local patterns from the traffic, which helps reduce the computational burden. 
The resulting feature representations are then fed into a Transformer encoder, which consists of two main components: a multi-head self-attention for capturing global dependencies, and a feed-forward network (FFN) for enhancing feature transformation and abstraction.

Due to the lack of inductive biases such as locality and translation invariance, which are inherent in CNN, the self-attention struggles to capture positional relationships between elements. To mitigate this limitation, we incorporate a learnable positional encoding $p\in \mathbb{R}^{M\times C}$, which is added to the feature map, resulting in a position-aware representation $z^\prime \in \mathbb{R}^{M\times C}$ as follows:

\begin{equation}
z^\prime=z + p
\end{equation}

\begin{figure}[!ht]
	\centering
	\includegraphics[width=0.99\linewidth]{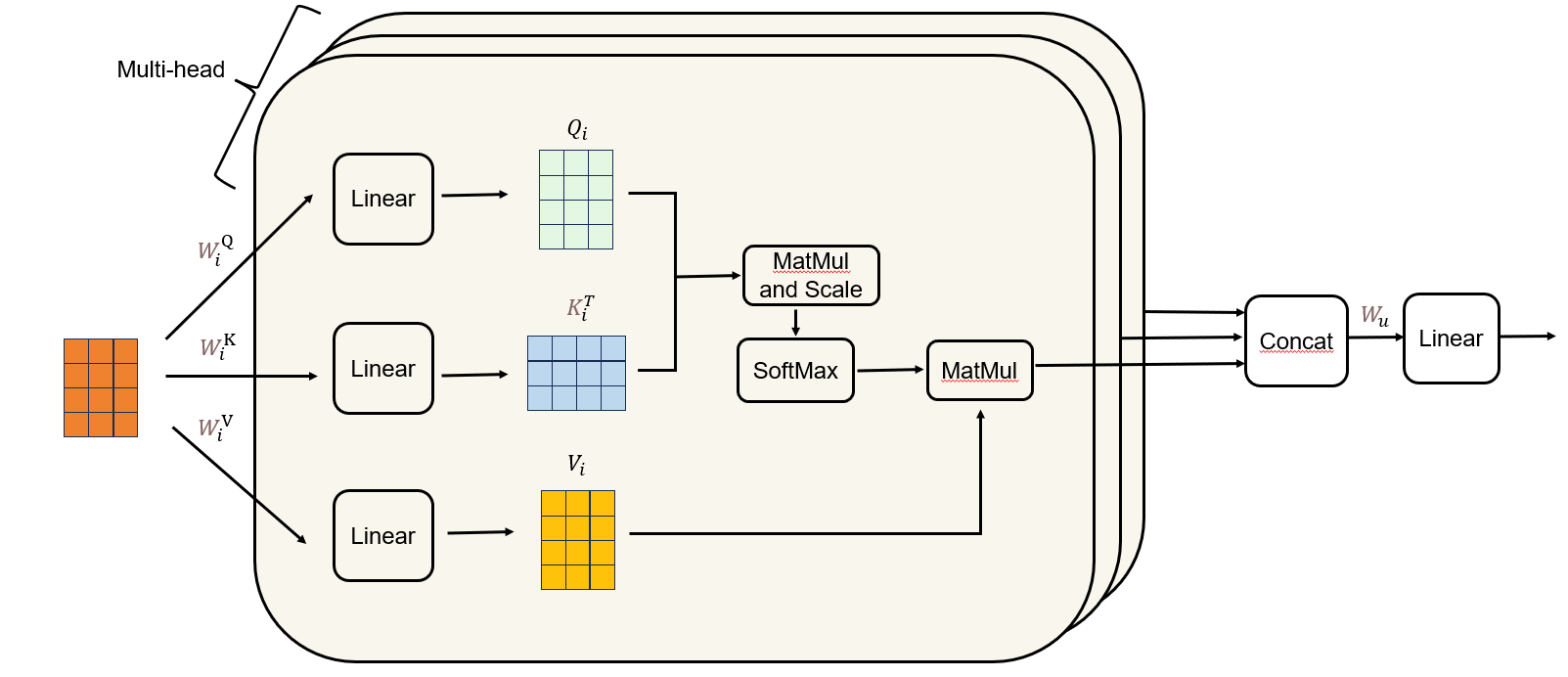}	
	\caption{Multi-head self-attention captures relationships between different positions in a sequence from multiple perspectives.} 
	\label{MHSA}%
\end{figure}

Fig.\ref{MHSA} illustrates the multi-head self-attention. The input feature map ${z^\prime}$ is first linearly projected into three matrices: the query $Q\in \mathbb{R}^{M\times C}$, key $K\in \mathbb{R}^{M\times C}$, and value $V\in \mathbb{R}^{M\times C}$. These are obtained using learnable weight matrices $W^Q\in \mathbb{R}^{C\times C}$, $W^K\in \mathbb{R}^{C\times C}$, $W^V\in \mathbb{R}^{C\times C}$ as follows:

\begin{equation}
    Q={z^\prime}W^Q,K={z^\prime}W^K,V={z^\prime}W^V,
\end{equation}

To model the relationships between different positions, we compute the scaled dot-product between $Q$ and $K$, and apply a softmax operation to obtain the attention weights. These weights are then used to aggregate the corresponding values in $V$, resulting in the final attention output. The full computation is given by:

\begin{equation}
    \mathrm{Attention}(Q,K,V)=\mathrm{softmax}(\frac{QK^T}{\sqrt{C}})V ,
\end{equation}

To reduce reliance on a single attention pattern and improve the model's representational capacity, we adopt a multi-head self-attention. In this framework, each attention head learns its own set of query
$Q_i\in \mathbb{R}^{M\times C_h}$, key $K_i\in \mathbb{R}^{M\times C_h}$, and value $V_i\in \mathbb{R}^{M\times C_h}$ matrices, which are derived by splitting the input along the channel dimension. Here, $C_h=C/h$ denotes the dimensionality of each head, where $h$ is the total number of attention heads. This allows the model to capture diverse types of dependencies and representations in parallel. The attention output for each head is computed as follows:

\begin{equation}
    head_i=\text{Attention}(Q_i,K_i,V_i) ,
\end{equation}

Subsequently, these outputs are concatenated along the channel dimension and projected using a linear transformation with a weight matrix $W_u \in \mathbb{R}^{hC_h\times C}$ to produce the aggregated output. To avoid over-fitting, layer normalization is applied across the feature dimensions for each sample, resulting in the intermediate representation $u\in \mathbb{R}^{M\times C}$. Finally, a residual connection adds the original input $z^\prime$ back $u$ to yield the final output $u^\prime \in \mathbb{R}^{M\times C}$. The computation is summarized as follows:

\begin{equation}
    u=\operatorname{Concat}(head_1,\ldots,head_h)W_u,
\end{equation}

\begin{equation}
    u^\prime=\mathrm{Layer}\mathrm{Norm}(u + z^\prime),
\end{equation}

The FFN layer primarily enhances the representation of each position through nonlinear transformations. Typically, the hidden dimension is expanded to four times the input dimension to increase the model's capacity.  The overall process is as follows.

\begin{equation}
    p=\sigma(u^\prime W_1)W_2,
\end{equation}

\begin{equation}
    q=\mathrm{Layer}\mathrm{Norm}(p + u^\prime),
\end{equation}

Here, $W_1 \in \mathbb{R}^{C\times C_f}$ and $W_2 \in \mathbb{R}^{C_f\times C}$ represent linear transformations, where $C_f$ denotes the expanded dimensionality, $\sigma$ is the nonlinear activation function, and $p \in \mathbb{R}^{M\times C}$ is the output after the nonlinear activation. A residual connection and layer normalization are applied at the end , resulting in the final output $q \in \mathbb{R}^{M\times C}$.

In the Transformer architecture, the encoder can be composed of multiple stacked layers to capture progressively abstract representations. Upon passing through these layers, we obtain the final feature representation $o \in \mathbb{R}^{M\times C}$.

\subsection{Residual Attention}

Since a single trace may contain multiple webpage categories, it is necessary to adaptively learn which segments correspond to which classes to achieve better discrimination. Given a feature matrix $o$,  we decompose it as $o_1,o_2,\ldots,o_{M}(o_j\in\mathbb{R}^{C})$, A fully connected layer with parameters $m_i \in\mathbb{R}^{C}$ is then used to project each segment into $W_n$ website classes. The class-specific attention score at position $j$ for category $i$ is computed as follows:

\begin{equation}
s_j^i=\frac{\exp(o_j^Tm_i)}{\sum_{j=1}^{M}\exp(o_j^Tm_i)},
\end{equation}

$s_j^i$ denotes the probability of the $i$-th class being present at position $j$. Consequently, the weighted aggregated score $v^i$ for the $i$-th class is calculated as follows:

\begin{equation}
v^i=\sum_{j=1}^{M}s_j^i{o}_j,
\end{equation}

In addition, a class-agnostic prediction score $g$ is derived using global average pooling, as shown below:

\begin{equation}
g=\frac{1}{M}\sum_{j=1}^{M}{o}_j,
\end{equation}

The two scores $g$ and $v^i$ are added to obtain the final prediction probability for the $i$-th class, denoted as the residual attention feature $r^i$, as computed below:

\begin{equation}
{r}^i=g+\lambda v^i,
\end{equation}

Finally, each prediction probability $r$ is compared with the ground-truth label $Y$ to compute the loss.

\section{Experiment}
\label{Experiment}

\subsection{Dataset}

We conduct experiments on the publicly available datasets, Oscar-1000 and Oscar-1001 \citep{zhao2024towards}, where each visited webpage is treated as a distinct class, and each trace may contain multiple webpages. These datasets are currently the largest multi-tab WPF benchmark, with the number of webpages per trace varying randomly,  reflecting realistic user browsing patterns. The dataset is divided into two scenarios: the closed-world setting and the open-world setting, which are described below:

\textbf{Oscar-1000.} In the closed-world setting, the dataset includes 115 websites selected from the Alexa Top 20,000 list. For each website, the homepage and 10 internal linked subpages are accessed via the Tor network. During each browsing session, from 1 to 5 pages are randomly selected and sequentially accessed, with intervals between webpages ranging from 3 to 10 seconds. This process yields 1,000 distinct webpage labels in total.

\textbf{Oscar-1001.} In the open-world setting, websites used in the closed-world are excluded from the Alexa Top 20,000 list, and the remaining homepages are retained as non-monitored classes. This results in 9,236 distinct websites treated as non-monitored pages. To simulate mixed monitored and non-monitored traffic, each session includes one non-monitored page, while the remaining pages, ranging from 2 to 5 pages per session, are selected from monitored classes. Each combination includes a unique non-monitored page.

\subsection{Experiment setup}
\label{Experiment setup}

All experiments are conducted using Python 3.8 and PyTorch 2.4. Model training and evaluation are performed on a single NVIDIA GeForce RTX 3090 GPU. The proposed ADWPF model is compared against five state-of-the-art baselines: DF \citep{sirinam2018deep}, BAPM \citep{guan2021bapm}, TMWF \citep{jin2023transformer}, ARES \citep{deng2023robust}, and Oscar \citep{zhao2024towards}. Among the baselines, DF is originally designed for single-tab WF. To enable multi-label classification, we replace its softmax output layer with a sigmoid activation function to constrain the outputs between 0 and 1, and substitute its loss with binary cross-entropy loss. For BAPM and TMWF, since they predict label indices instead of probability scores and rely on prior knowledge of the maximum number of pages, we unify their multiple prediction heads by averaging their outputs. These models are also trained using binary cross-entropy loss to ensure consistency in evaluation metrics. Models such as Tik-Tok \citep{rahman2019tik} and Var-CNN \citep{bhat2018var} are excluded from our comparison as they incorporate additional temporal information, which is not considered in the current experimental setup.

\begin{itemize}
  \item \textbf{DF} \citep{sirinam2018deep} employs a CNN architecture to extract direction features from traces and integrates regularization techniques such as batch normalization and dropout to prevent overfitting, achieving significant results in single-tab WF.
  \item \textbf{BAPM} \citep{guan2021bapm} adopts a self-attention mechanism to enable end-to-end identification of multiple websites without explicit segmentation, generating tab-aware representations while preserving the integrity of the trace.
  \item \textbf{TMWF} \citep{jin2023transformer}  utilizes a Transformer encoder to learn global representations of traffic and employs a tab query mechanism for each website category, which is then fed into the decoder to fuse with traffic features.
  \item \textbf{ARES} \citep{deng2023robust} extracts local patterns using a sliding window and CNN, and analyzes their relationships via multi-head top-k self-attention.
  \item \textbf{Oscar} \citep{zhao2024towards} adopts a metric learning paradigm to capture subtle differences between webpages. It transforms direction features into an embedding space and computes label scores based on both proxy-based and sample-based k-NN results.
\end{itemize}

\textbf{Metrics.} We employ three multi-label evaluation metrics: mean Average Precision (mAP) \citep{zhu2021residual}, Recall@k \citep{chen2017task}, and AP@k \citep{zhang2014lift}. These metrics are independent of thresholds, offering a more intuitive and comprehensive assessment of model performance. Recall@k measures whether ground-truth websites are included within the top-k predictions. Specifically, given a ground-truth label set $Y_i$ for a sample, the set of positive labels is denoted as $Y_{i}^{+}$, and the top-k predicted labels with the highest probabilities are represented as $\hat{Y}^k_i$. Recall@k is computed as follows:

\begin{equation}
Recall@k=\frac{|Y_i\cap{\hat{Y}^k_i}|}{|Y_{i}^{+}|}
\end{equation}

AP@k is used to evaluate the ranking accuracy of a sample by considering only the top-k retrieved results. It begins with the computation of P@k, which is defined by the following formula:

\begin{equation}
P@k=\frac{|Y_i\cap{\hat{Y}^k_i}|}{k}
\end{equation}

Subsequently, only the scores of relevant items, denoted as $Y_{i}^{+}$, are considered; incorrect predictions are excluded to avoid contributions. Here, $r(i)$ denotes whether the $i$-th item is relevant (1 if relevant, 0 otherwise). The AP@k calculation is expressed as follows:

\begin{equation}
AP@k=\frac{1}{\min(|Y_{i}^{+}|,k)}\sum_{i=1}^kP@i\cdot\mathrm{r}(i)
\end{equation}

\textbf{Parameters.} The detailed model parameters are shown in Table \ref{parameters}. In all experiments, the input sequence length is fixed at 10,000. After processing through the backbone, the feature representation has a shape of $M \times C = 16 \times 640 $.  In the data augmentation stage, the cropping threshold $\varphi_c$ is randomly sampled between 0.4 and 0.6, while the masking threshold $\varphi_m$ is sampled between 0.2 and 0.5. During cropping, we additionally trim 1000 packets on both the left and right sides to prevent excessive information loss. For the residual attention module, $\lambda$ is set to 0.3. The model is trained for 100 epochs using the Adam optimizer with a learning rate of $1e^{-4}$ and a batch size of 64.

\begin{table}[!htp]
\centering
\captionsetup{margin=2.5cm} 
\caption{Parameter settings for ADWPF.}
\begin{adjustbox}{max width=\linewidth}
\label{parameters}
\begin{tabular}{ccc}
\toprule
\textbf{Module Part}                  & \textbf{Design Details}           & \textbf{Value}               \\
\hline
\multirow{5}{*}{Backbone}    & Input dimension          & 10000           \\
                             & Number  of filters      & {[}64,160,320,640{]} \\
                             & Kernel   size            & {[}3,3,3,3{]}       \\
                             & Pool   size              & {[}9,9,9,9{]}       \\
                             & Output dimension          & (16,640) \\
\hline

\multirow{2}{*}{Transformer} & Number of layers         & 4     \\    
                             & Number of heads            & 8                   \\

\bottomrule
\end{tabular}
\end{adjustbox}
\end{table}

\subsection{Experimental results and analysis}

\subsubsection{Closed-world}

\begin{table}[h!]
\centering
\caption{Performance comparison of prior arts in the closed-world scenario (\%).}
\begin{adjustbox}{max width=\linewidth}
\label{closed-world-table}
\begin{threeparttable}
\begin{tabular}{lcccccc}
\toprule
\textbf{Attacks} & DF \citep{sirinam2018deep} & BAPM \citep{guan2021bapm} & TMWF \citep{jin2023transformer} & ARES \citep{deng2023robust} & Oscar \citep{zhao2024towards} \tnote{1} & ADWPF \\
\midrule
\textbf{Recall@5} & 40.09 & 33.90 & 43.24 & 53.77 & 48.72 & \textbf{63.85} \\
\textbf{AP@5}     & 29.65 & 25.45 & 34.33 & 44.42 & 36.87 & \textbf{55.67} \\
\textbf{mAP}      & 20.02 & 16.22 & 25.93 & 37.09 & -- & \textbf{50.54} \\
\bottomrule
\end{tabular}
\begin{tablenotes}
    \item[1] Oscar simultaneously employs both proxy-based k-NN and sample-based k-NN for prediction, resulting in the summation of two probabilities, which makes it unsuitable for mAP computation. 
\end{tablenotes}

\end{threeparttable}

\end{adjustbox}
\end{table}

We begin by evaluating the performance of ADWPF and several baseline methods in the closed-world setting, where users only access webpages from the monitored set. The dataset is divided into training, validation, and test sets in a ratio of 8:1:1. Evaluation metrics include mean Average Precision (mAP), Recall@k (where k = 5, 10, 15, 20), and AP@k (where k = 1, 2, 3, 4, 5). As shown in Table \ref{closed-world-table}, ADWPF achieves the best performance in terms of Recall@5, AP@5, and mAP. Specifically, its Recall@5 surpasses that of DF, BAPM, TMWF, ARES, and Oscar by 23.76\%, 29.95\%, 20.61\%, 10.08\%, and 15.13\%, respectively. These results demonstrates that ADWPF accurately and effectively identifies webpages in the closed-world scenario.

\begin{figure}[!ht]
	\centering 
	\includegraphics[width=0.99\textwidth]{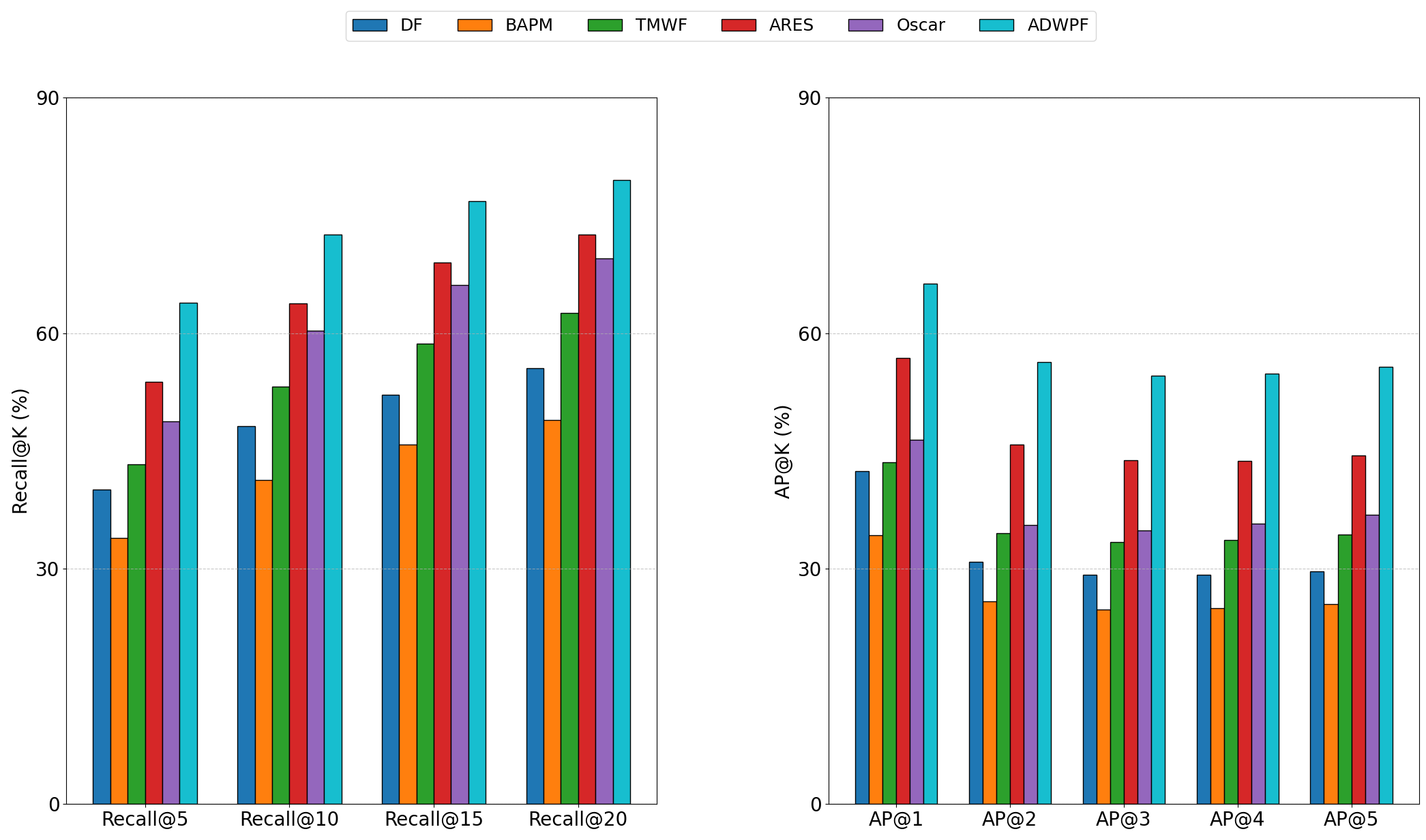}	
	\caption{Comparison of model performance in terms of Recall@k and AP@k at different values of k in the closed-world setting.}
	\label{closed-world-fig}%
\end{figure}

As shown in Fig.\ref{closed-world-fig}, we increase the value of k to observe the model's performance. ADWPF's Recall@k consistently increases from 5 to 20 and remains significantly higher than those of all baselines. Although AP@k exhibits some fluctuations, it still outperforms all other models by at least 10\%.

Although ARES is currently the strongest baseline, its performance deteriorates when dealing with large-scale webpage traffic. In contrast, ADWPF leverages a more powerful feature extractor to ensure the richness of extracted local features. Moreover, it incorporates attention-based enhancement mechanisms to help the model capture more discriminative and previously overlooked features. The residual attention module further facilitates the model’s ability to focus on features across spatial locations.  Furthermore, Oscar falls behind ARES, largely due to the suboptimal feature space learned by its feature extractor, which adversely affects downstream webpage identification.

\subsubsection{Open-world}

\begin{table}[h!]
\centering
\caption{Performance comparison of prior arts in the open-world scenario (\%).}
\begin{adjustbox}{max width=\linewidth}
\label{open-world-table}
\begin{threeparttable}
\begin{tabular}{lcccccc}
\toprule
\textbf{Attacks} & DF \citep{sirinam2018deep} & BAPM \citep{guan2021bapm} & TMWF \citep{jin2023transformer} & ARES \citep{deng2023robust} & Oscar \citep{zhao2024towards} \tnote{1} & ADWPF \\
\midrule
\textbf{Recall@5} & 38.68 & 31.19 & 40.09 & 49.71 & 44.99 & \textbf{58.87} \\
\textbf{AP@5}     & 28.49 & 22.97 & 31.14 & 40.38 & 33.63 & \textbf{50.49} \\
\textbf{mAP}      & 18.37 & 13.93 & 22.68 & 32.82 & -- & \textbf{44.12} \\
\bottomrule
\end{tabular}
\begin{tablenotes}
    \item[1] Oscar simultaneously employs both proxy-based k-NN and sample-based k-NN for prediction, resulting in the summation of two probabilities, which makes it unsuitable for mAP computation. 
\end{tablenotes}

\end{threeparttable}
\end{adjustbox}
\end{table}

\begin{figure}[!ht]
	\centering 
	\includegraphics[width=0.99\textwidth]{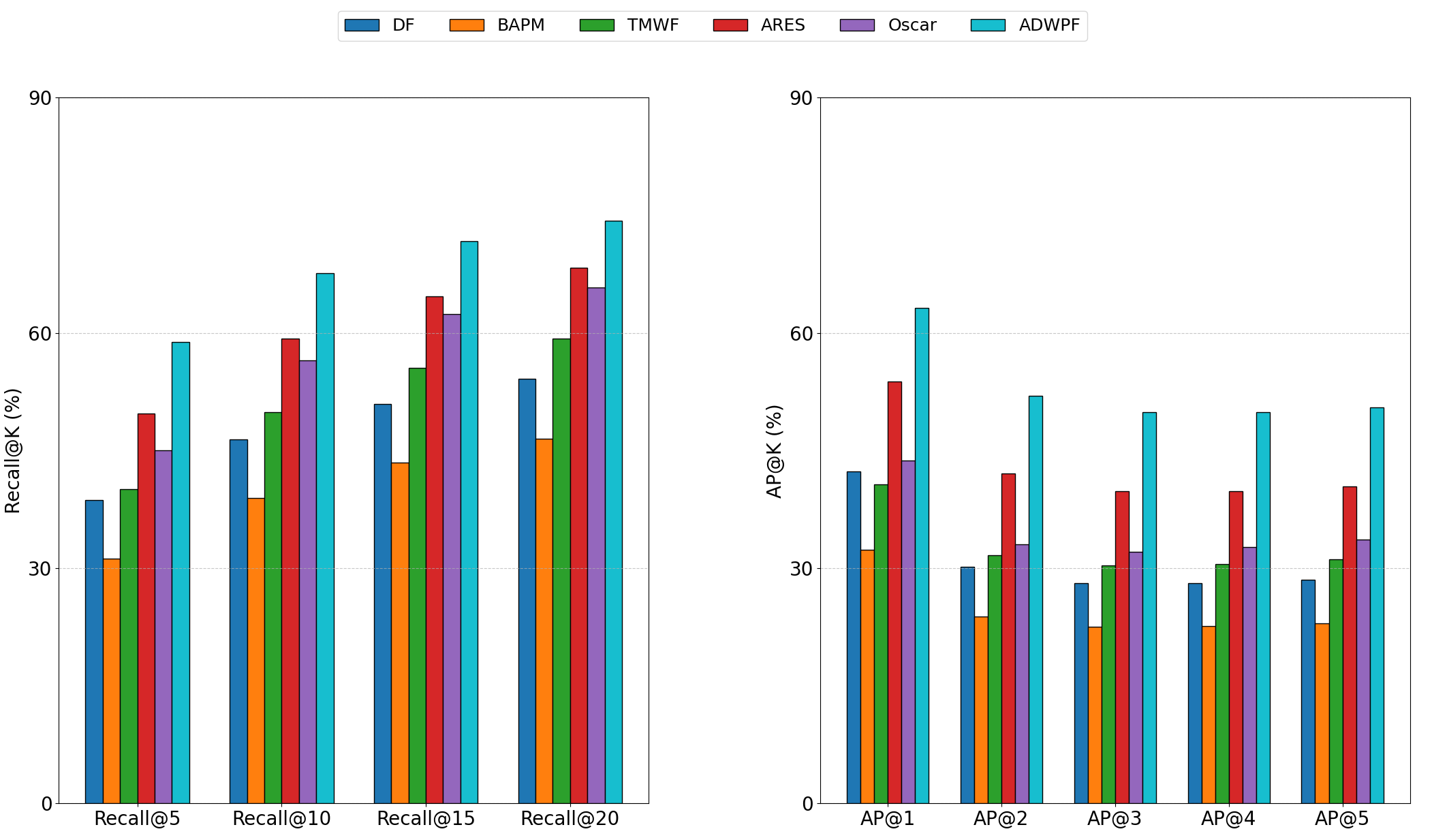}	
	\caption{Comparison of model performance in terms of Recall@k and AP@k at different values of k in the open-world setting.}
	\label{open-world-fig}%
\end{figure}

In the open-world setting, to better evaluate the model's generalization capability, it must not only predict webpages within the monitored set but also identify those that do not belong to it. We treat all non-monitored webpages as a single class, while the configuration for monitored pages remains the same as in the closed-world scenario. The dataset is still split into training, validation, and test sets with a ratio of 8:1:1.

As shown in Table \ref{open-world-table}, in the open-world setting, ADWPF achieves a Recall@5 of 58.87\% and an AP@5 of 50.49\%, outperforming the best baseline, which reach only 49.71\% and 40.38\%, respectively. Even in the presence of a large number of unmonitored webpages, ADWPF still demonstrates superior accuracy compared to other baselines. Similarly, we increase the value of k to better observe how model performance scales, as illustrated in Fig.\ref{open-world-fig}.
ADWPF achieves a Recall@20 of approximately 74\%, outperforming ARES by about 5\%, and consistently surpasses all other models across all k values. Notably, compared to the closed-world results, all models exhibit a slight performance degradation, primarily due to the substantial content diversity present in the non-monitored set. This diversity makes it difficult for models to fully capture all possible webpage characteristics.

Nevertheless, ADWPF demonstrates strong generalization even in the presence of more complex data distributions. Its attention-based module effectively boosts its ability to generalize, while the residual attention mechanism enables the model to adaptively localize non-monitored webpages. These results demonstrate that attention mechanisms in ADWPF confer greater flexibility when operating in complex and dynamic network environments.

\subsubsection{Various scales of webpages}

\begin{table}[!ht]
\centering
\captionsetup{margin=1.5cm}
\caption{Number of training, validation, and test samples under different scales of monitored webpages. (\%).}
\begin{adjustbox}{max width=\linewidth}
\label{statistics of samples}
\begin{tabular}{ccccc}
\toprule
\textbf{Number of webpages} & \textbf{Training set}  & \textbf{Validation set}  & \textbf{Test set} & \textbf{Total} \\
\midrule
700 & 56,711 & 7,089 & 7,089 & 70,889 \\
800 & 59,746 & 7,468 & 7,469 & 74,683 \\
900 & 61,909 & 7,739 & 7,739 & 77,387 \\
1,000 & 65,027 & 8,128 & 8,129 & 81,284 \\
\bottomrule
\end{tabular}
\end{adjustbox}
\end{table}

\begin{figure}[!htp]
	\centering 
	\includegraphics[width=0.99\textwidth]{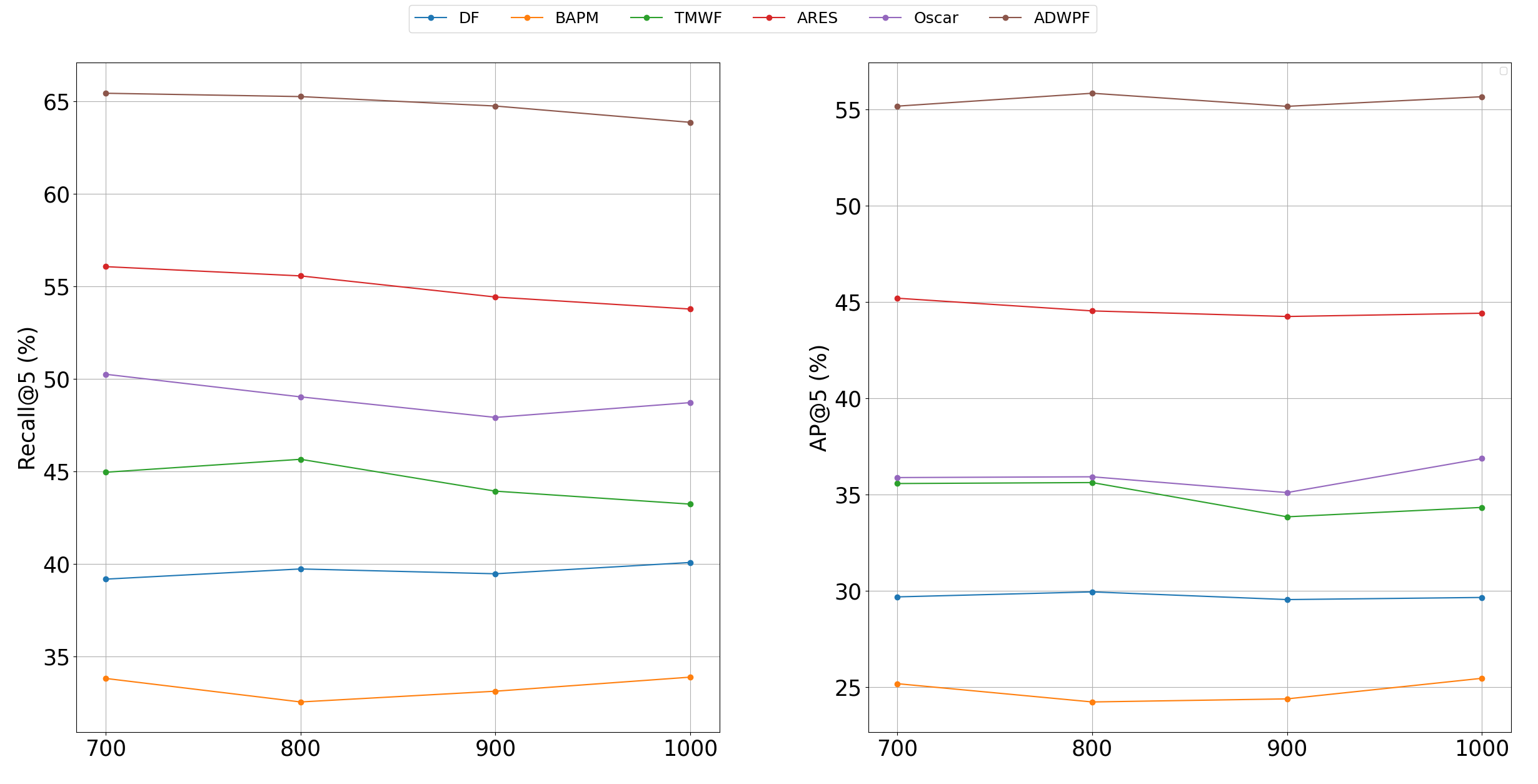}	
	\caption{Comparison of Recall@5 and AP@5 across different monitored webpage scales.}
	\label{world-size-fig}%
\end{figure}

We further evaluate the model's performance under varying scales of monitored webpages, using datasets containing 700, 800, 900, and 1000 monitored pages, respectively. The construction of these datasets is as follows: we randomly sample different numbers of webpages from the closed-world dataset and extract the corresponding traffic samples. Only traces whose labels overlap with the selected webpages are retained. A sample that contains labels not present in the selected pages is discarded. For webpages of different scales, we still divide the dataset into training, validation, and test sets in an 8:1:1 ratio, with the specific sample numbers shown in Table \ref{statistics of samples}. We assess model performance using Recall@5 and AP@5, as illustrated in Fig.\ref{world-size-fig}.

Experimental results show that as the number of monitored webpages increases, the Recall@5 of ADWPF exhibits only a slight decline, consistently remaining above 60\%, and AP@5 fluctuates around 55\%. Both metrics are significantly higher than those of all baselines. This indicates that the model’s performance is not substantially affected by the scale of the monitored set. The residual attention mechanism effectively enables the model to adaptively locate the relevant features of target webpages even among hundreds or thousands of candidates, allowing ADWPF to maintain strong performance across varying configurations.

\subsubsection{Ablation study}

\begin{table}[!ht]
\centering
\captionsetup{margin=2.5cm}
\caption{Ablation study results on ADWPF (\%).}
\begin{adjustbox}{max width=\linewidth}
\label{ablation}
\begin{threeparttable}
\begin{tabular}{c|c|c|c|c|c|c}
\toprule
\textbf{RA} \tnote{1} & \textbf{AC} \tnote{2} & \textbf{AM} \tnote{3} & \textbf{RAtt} \tnote{4} & \textbf{Recall@5} & \textbf{AP@5} & \textbf{mAP} \\
\midrule
& & & & 58.19 & 48.72 & 36.25 \\
\checkmark & & & & 57.55 & 47.69 & 35.20 \\ 
& \checkmark && & 60.94 & 51.02 & 38.12 \\  
& & \checkmark & & 63.90 & 53.71 & 41.03 \\  
& \checkmark & \checkmark & & 64.77 & 54.47 & 41.88 \\  
& \checkmark &  \checkmark  & \checkmark & \textbf{65.42} & \textbf{55.18} & \textbf{42.70} \\ 
\bottomrule
\end{tabular}

\begin{tablenotes}
    \item[1] RA represents the random augmentation. 
    \item[2] AC represents the attention cropping. 
    \item[3] AM represents the attention masking. 
    \item[4] RAtt represents the residual attention.
\end{tablenotes}

\end{threeparttable}
\end{adjustbox}
\end{table}

We then conduct ablation experiments to evaluate the contributions of attention augmentation and residual attention to the model’s performance. In addition, we introduce a random augmentation strategy as a baseline. This method involves randomly cropping and masking directions from the training set, and then mixing the modified data with the original training data to create a simple augmented dataset. We conducted the ablation study on a dataset consisting of 700 monitored webpages.

As shown in Table \ref{ablation}, when using only the CNN and Transformer architecture, the model achieves an mAP of 36.25\%. Surprisingly, applying random augmentation results in a performance drop, with mAP decreasing by 1.05\%. This suggests that random augmentation may lead to information loss in multi-tab traffic, thereby impairing the model’s generalization ability. 
We further compare different attention augmentation strategies. Attention cropping improves model’s mAP by 1.87\%, and attention masking results in a 4.78\% increase. 
The superior performance of attention masking alone suggests that many regions of the traffic are previously underutilized by the model. When both attention cropping and masking are combined, the mAP improves by 5.63\%, highlighting that targeted enhancement is substantially more effective than random augmentation.
Finally, when residual attention is incorporated, the model achieves the best performance, with Recall@5, AP@5, and mAP reaching 65.42\%, 55.18\%, and 42.70\%, respectively. Residual attention encourages the model to discover discriminative features across different spatial positions in multi-tab traffic, leading to more robust and accurate predictions.

\begin{table}[!ht]
\centering
\captionsetup{margin=1.8cm}
\caption{Number of samples under different tab settings with 700 monitored webpage classes. (\%).}
\begin{adjustbox}{max width=\linewidth}
\label{statistics of tabs}
\begin{tabular}{lcccccc}
\toprule
\textbf{Number of tabs} & 1-tab & 2-tab & 3-tab & 4-tab & 5-tab & Total \\
\midrule
\textbf{Number of samples} & 2,913 & 2,517 & 1,145 & 437 & 77 & 7,089 \\
\bottomrule
\end{tabular}
\end{adjustbox}
\end{table}

Another important consideration is whether data augmentation remains effective as the number of tabs increases. In large-scale scenarios where multiple pages are loaded concurrently, attention cropping and masking may inadvertently discard important information. Table \ref{statistics of tabs} shows the label statistics of the test set under the 700 monitored webpage classes. To investigate this, we evaluate the model’s performance under varying tab counts with 700 monitored pages, as shown in Table \ref{effect of tabs}.

\begin{table}[!ht]
\centering
\caption{Effect of different attention augmentation strategies across the number of tabs (\%).}
\begin{adjustbox}{max width=\linewidth}
\label{effect of tabs}
\begin{threeparttable}
\begin{tabular}{ccc|ccc|ccc|ccc|ccc|ccc|}
\toprule
\multirow{2}{*}{\textbf{RA} \tnote{1}} & \multirow{2}{*}{\textbf{AC} \tnote{2}}& \multirow{2}{*}{\textbf{AM} \tnote{3}} & \multicolumn{3}{c}{\textbf{1-tab}}     & \multicolumn{3}{c}{\textbf{2-tab}}     & \multicolumn{3}{c}{\textbf{3-tab}}     & \multicolumn{3}{c} 
{\textbf{4-tab}} & \multicolumn{3}{c} 
{\textbf{5-tab}}   \\ \cmidrule(lr){4-6} \cmidrule(lr){7-9} \cmidrule(lr){10-12} \cmidrule(lr){13-15} \cmidrule(lr){16-18}
&   &    &      \textbf{AP@1}     & \textbf{Recall@5}    & \textbf{mAP}   & \textbf{AP@2 }    & \textbf{Recall@5}     & \textbf{mAP}    & \textbf{AP@3}   & \textbf{Recall@5}     & \textbf{mAP}     & \textbf{AP@4}    & \textbf{Recall@5}   & \textbf{mAP}     & \textbf{AP@5}     & \textbf{Recall@5}    & \textbf{mAP}   \\
                       \midrule
& &  & 47.27 & 65.74 & 39.45 & 41.73 & 56.50 & 39.67 & 37.05 & 47.98 & 38.39 & 39.34 & 46.97 & 39.38 & 39.20 & 43.64 & 16.72 \\

\checkmark & &  & 46.34 & 66.15 & 39.93 & 40.82 & 55.88 & 38.99 & 35.18 & 46.03 & 36.00 & 35.67 & 43.42 & 36.38 & 34.30 & 38.96 & 16.84 \\

&\checkmark &  & 49.06 & 68.55 & 41.19 & 43.83 & 59.50 & 41.63 & 38.71 & 50.22 & 39.41 & 41.34 & 48.86 & 40.44 & 44.35 & 48.31 & 18.20 \\

&& \checkmark  & 50.84 & 71.47 & 43.32 & 46.64 & 62.02 & 44.53 & 42.31 & 54.50 & 42.55 & 43.31 & 51.49 & 42.72 & 44.16 & 49.09 & 18.81 \\

&\checkmark & \checkmark  & \textbf{51.49} & \textbf{72.09} & \textbf{43.95} & \textbf{47.58} & \textbf{63.77} & \textbf{45.36} & \textbf{42.47} & \textbf{54.03} & \textbf{43.26} & \textbf{44.73} & \textbf{51.95} & \textbf{44.02} & \textbf{47.75} & \textbf{52.99} & \textbf{19.59} \\

\bottomrule
\end{tabular}

\begin{tablenotes}
    \item[1] RA represents the random augmentation. 
    \item[2] AC represents the attention cropping. 
    \item[3] AM represents the attention masking. 
\end{tablenotes}

\end{threeparttable}
\end{adjustbox}
\end{table}

It is evident that as the number of tabs increases, all evaluation metrics experience a significant decline. Comparing the 1-tab and 5-tab settings, the model without any augmentation strategy shows a decrease of 22.1\% in Recall@5 and a 22.73\% in mAP. This suggests that a greater number of pages makes the traffic patterns more complex, primarily due to the overlapping content between webpages. 

A vertical comparison of different augmentation strategies reveals that random augmentation has virtually no positive impact across various tab settings. Under the 1-tab setting, it only leads to improvements of 0.41\% in Recall@5 and 0.48\% in mAP, while in other settings, performance mostly declines. This indicates that such a method fails to enhance the model’s generalization capability. However, attention cropping and attention masking outperform the no-augmentation baseline across all tab settings. Under the most challenging 5-tab setting, AP@5, Recall@5, and mAP improved by 8.55\%, 9.35\%, and 2.87\%, respectively. In contrast, random cropping and random masking are more likely to cause the loss of key regions in multi-tab settings. These findings further indicate that enabling the model to learn augmentation strategies adaptively during training more effectively enhances its performance.

\subsubsection{Analysis of augmented data}

\begin{figure}[H]
  \centering
  \begin{subfigure}{\linewidth}
    \centering
    \begin{subfigure}{0.48\linewidth}
      \centering
      \includegraphics[width=\linewidth]{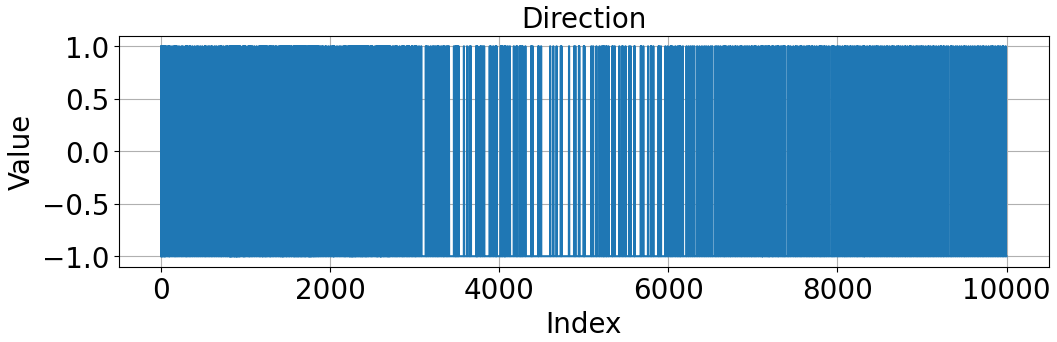}
    \end{subfigure}
    \hfill
    \begin{subfigure}{0.48\linewidth}
      \centering
      \includegraphics[width=\linewidth]{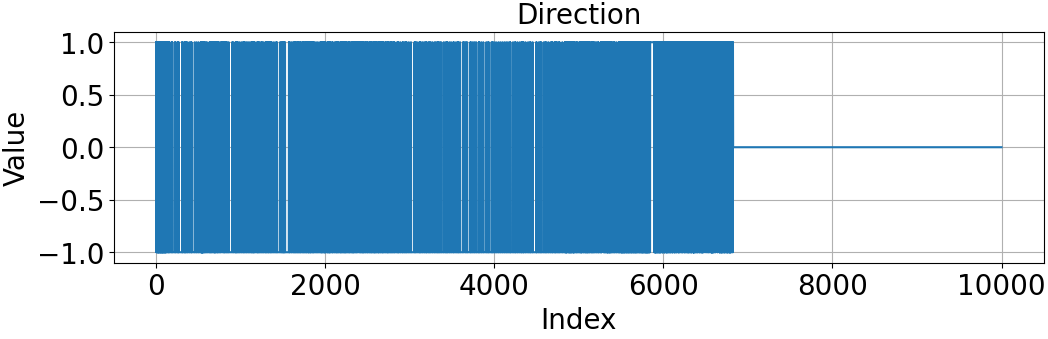}
    \end{subfigure}
    \caption{Original samples.}
  \end{subfigure}

  \begin{subfigure}{\linewidth}
    \centering
    \begin{subfigure}{0.48\linewidth}
      \centering
      \includegraphics[width=\linewidth]{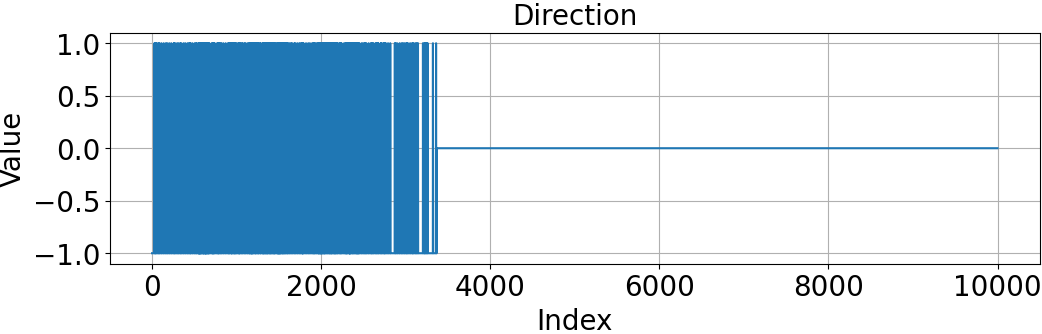}
    \end{subfigure}
    \hfill
    \begin{subfigure}{0.48\linewidth}
      \centering
      \includegraphics[width=\linewidth]{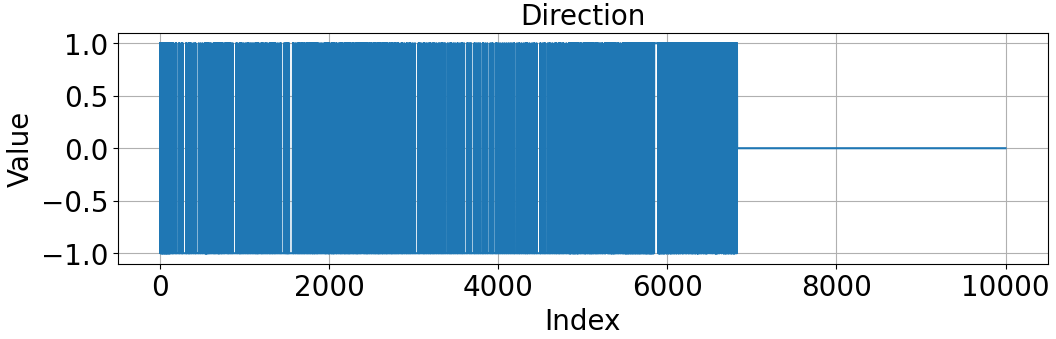}
    \end{subfigure}
    \caption{Attention cropping samples.}
  \end{subfigure}

  \begin{subfigure}{\linewidth}
    \centering
    \begin{subfigure}{0.48\linewidth}
      \centering
      \includegraphics[width=\linewidth]{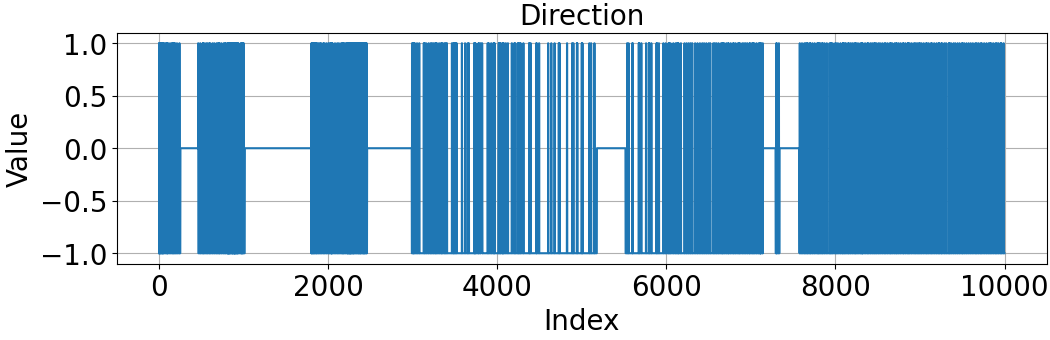}
    \end{subfigure}
    \hfill
    \begin{subfigure}{0.48\linewidth}
      \centering
      \includegraphics[width=\linewidth]{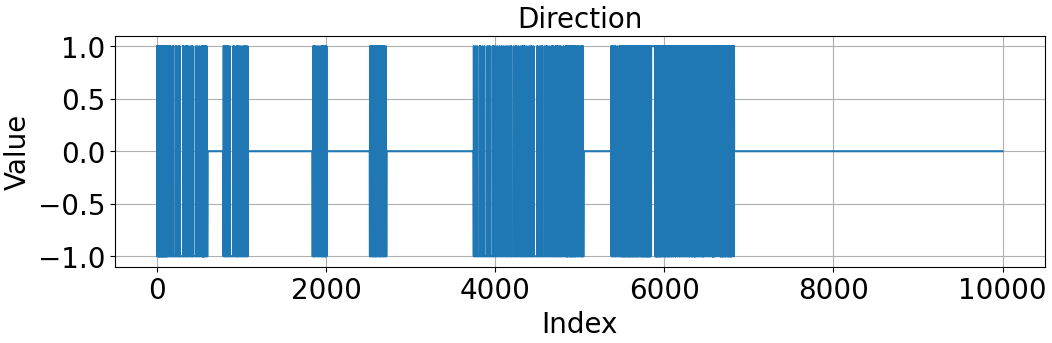}
    \end{subfigure}
    \caption{Attention masking samples.}
  \end{subfigure}

  \caption{The results of applying attention cropping and attention masking to the original sample.}
  \label{analysis-augmented-data}
\end{figure}

To better understand how attention augmentation functions, we visualized two samples processed by attention cropping and attention masking, as shown in Fig.\ref{analysis-augmented-data}. The blue regions represent areas where the traffic direction fluctuates rapidly between +1 and -1, while the white regions indicate bursts in which the direction remains consistently at either +1 or -1. As shown in the second row, the first sample experiences a significant change in sequence length after cropping, whereas the second sample shows minimal alteration. Since multi-tab traffic corresponds to multiple, spatially distinct regions, the regions of interest may become overly dispersed. 
To mitigate this region, we crop an additional 1000 cells on both sides during the process, reducing the risk of the model becoming overconfident in predicting a category. After masking, many parts of the input are set to zero, while the remaining blue regions align with traffic segments that the model typically overlooks. These two augmentation operations help the model learn features that differ from those in the original samples.

\section{Discussion}
\label{Discussion}

The results of ADWPF demonstrate its strong performance in multi-tab WPF attacks. Models such as DF and BAPM, which rely on shallow feature extractors, struggle to capture semantically rich representations. In contrast, TMWF and ARES utilize CNN to extract local patterns and incorporate multi-head self-attention to model global relationships , achieving better identification performance. However, their generalization capabilities are still constrained by the absence of effective data augmentation strategies. Oscar departs from the traditional end-to-end learning paradigm by employing metric learning to map direction features into a feature space, where both sample-to-sample and sample-to-proxy relationships are computed. Despite this innovative framework, Oscar exhibits inferior performance in both prediction accuracy and computational efficiency compared to end-to-end models like ARES and ADWPF. Moreover, ablation studies confirm that, compared to random augmentation, targeted attention augmentation can substantially improve model performance without relying on expert knowledge. However, there remain areas in ADWPF that require further improvement.

\textbf{Large Models and Training Overhead.} Current multi-tab WF attacks often rely on large models to capture rich feature representations. Although Oscar reduces the model size through metric learning, its inference speed is slower compared to end-to-end models. Additionally, both approaches result in significant training overhead. For instance, Oscar and ADWPF require approximately 12 hours of training on a single GPU. While techniques such as pre-training and fine-tuning can accelerate domain adaptation for large models, they remain impractical for deployment on resource-constrained edge devices. Therefore, future research will investigate lightweight model architectures and compression strategies that maintain fast response without incurring significant accuracy degradation.

\textbf{Subverting WPF defenses.} All experiments conducted in this study used datasets without any defenses. However, several prior works proposed various defenses to counter website fingerprinting attacks. For example, WTF-PAD \citep{juarez2016toward} and FRONT \citep{gong2020zero} employ packet padding techniques to obfuscate traffic features, while TrafficSilver \citep{de2020trafficsliver} hinders the collection of a complete trace by fragmenting traffic into multiple segments. Blanket \citep{nasr2021defeating} generates adversarial perturbations to disrupt model predictions. In future work, we aim to evaluate the robustness of WPF attacks in the presence of such defenses and explore potential countermeasures.

\textbf{Identifying Heavily Overlapped Traffic.} Current multi-tab traffic datasets typically contain no more than six webpages per sample. However, in real-world scenarios, users often access a dozen or more pages simultaneously, resulting in a significant increase in traffic overlap and interference. Existing methods generally train models on datasets with a limited number of webpages per sample, which restricts the model's ability to generalize to situations with greater numbers of pages. This limitation undermines the model's performance when confronted with more complex traffic patterns. In future work, we aim to address this limitation by constructing more complex datasets that better reflect real-world traffic or by integrating techniques that enhance identification in scenarios with numerous overlapping pages, enabling the model to adapt to a wider range of traffic scenarios.

\section{Conclusion}
\label{Conclusion}

This paper tackles the challenge of large-scale, multi-tab WPF attacks. 
Prior approaches largely focused on identifying website homepages, often neglecting the more realistic scenario in which users navigate across multiple subpages.
To address this limitation and support end-to-end WPF under complex conditions, we propose a novel attention-driven framework, ADWPF, which leverages attention mechanisms to enable the model to adaptively capture and differentiate the subtle variations between webpages, thereby enhancing its ability to identify fine-grained traffic patterns.

ADWPF leverages attention maps to identify highly discriminative regions and applies two complementary augmentation strategies: attention cropping, which focuses the model on salient regions, and attention masking, which encourages the exploration of overlooked areas.  To model complex dependencies across different parts of the traffic sequence, ADWPF employs self-attention for global context modeling. Finally, a residual attention module at the output layer facilitates the alignment of class-specific features with their corresponding positions, thereby improving multi-label classification performance.

We extensively evaluate ADWPF in both closed-world and open-world scenarios. Remarkably, under a realistic setting involving 1,000 website categories, our model achieves 50.54\% mAP, demonstrating strong generalization capability to complex, large-scale conditions. We hope this work provides new insights and tools to advance the development of WPF in real-world environments.

\section*{Acknowledgements}
This work was supported in part by the National Key Research and Development Program of China (Grant No. 2023YFB3106700) under the Young Scientists Program,  in part by Natural Science Foundation of Jiangsu Province  (Grant No. SBK2023041256), and in part by the National Natural Science Foundation of China (Grant No. 62302097).

\bibliographystyle{elsarticle-harv} 
\bibliography{reference-template}



\end{document}